\documentclass{article}
\usepackage{amssymb}

\usepackage{graphicx}
\usepackage{amsmath}

\newtheorem{theorem}{Theorem}

\input{tcilatex}

\begin{document}

\title{Quantum Stochastic Generators}
\author{John Gough \\
School of Computing and Informatics,\\
Nottingham Trent University,\\
Nottingham, NG1 4BU, United Kingdom}
\date{}
\maketitle

\begin{abstract}
We discuss stochastic derivations, stochastic Hamiltonians and the flows
that they generate, algebraic fluctuaion-dissipation theorems, etc., in a
language common to both classical and quantum algebras. It is convenient to
define distinct notions of time-ordered exponentials to take account of the
\ breakdown of the Leibniz rule in the It\^{o} calculus. We introduce a
notion of quantum Stratonovich calculus and show how it relates to
Stratonovich-Dyson time ordered exponentials. We then use it to demonstrate
a natural way to add stochastic derivations.
\end{abstract}

\section{Introduction}

Symmetries play a considerable role in Mathematical Physics, particularly
when determining dynamical flows with required invariant properties. Much of
our insight comes from being able to go from the infinitesimal generators
(having desirable symmetry features) to the flows themselves. This intuition
requires, to a large extent, the generators to be derivations. When we
consider stochastic flows, the It\^{o} calculus looses the Leibniz rule and
with it much of our physical intuition. This is why the Stratonovich
calculus\ \cite{S}\ is usually preferred by physicists. In addition, the
Wong-Zakai approximation theory tells us that we can approximate a
stochastic flow with a differentiable random flow: if the approximations
have some symmetry property then the limit Stratonovich equations will
retain this property before it is typically lost in the conversion to
It\^{o} form.

The aim of this article is describe algebraic notions such as derivations,
their stochastic analogues, dissipation, etc., in a common language that
applies to classical and quantum systems. Whereas there have been many
attempts to construct general classical analogues to quantum stochastic
flows, the spirit of the quantum flows is arguably best captured by
stochastic flows on symplectic manifolds preserving the Poisson brackets.
This was investigated by Sinha \cite{Sinha} and we develop somewhat the
algebraic similarities. Given the suitability of C*-algebras for modelling
quantum mechanical variables it is natural to consider algebras of functions
on Poisson manifolds as an intermediary between the commutative and
noncommutative cases. This view is strengthened considerably by the deep
analogies known to exist between Poisson and operator algebras \cite
{Weinstein}, \cite{Wmod}. It is natural to return to these analogies when
considering stochastic flows describing irreversible dynamical evolutions.

One of the most notable omissions from quantum probability has been a
general theory of a quantum Stratonovich calculus. This is largely due to
the bias, already prevalent in classical probability, towards the It\^{o}
calculus on the grounds that of martingale stability. Nevertheless, in
problems, such as stochastic process on manifolds \cite{Meyer} \cite{Emery},
the Stratonovich calculus is used in on entirely equivalent footing as the
It\^{o} calculus and, because it does maintain the Leibniz rule, allows one
to see the underlying differential geometric structure. Motivated to develop
these idea for algebras of quantum observables, the author considered
Stratonovich integral for quantum diffusions \cite{JGJMP}, however, in the
later development of this ideas, much of the physical insight came from
limit theorems - weak coupling and low density - and this lead to a
formulation in terms of quantum white noises: see \cite{G}, \cite{ALV}.
Formally, the familiar features, like a Hamiltonian nature for the flow, are
evident in the Weyl ordered form (the analogue of Stratonovich version)
while the Wick ordering distorts these features and produced the analogue of
the It\^{o} version. While quantum white noises offer the most intuitive
approach, and should amenable to a rigorous treatment in some extension of
the Hida theory, they have not found favour with either the quantum
probability or the mathematical physics community. However, it is possible
to\ give a sufficient account of events from within the quantum It\^{o}
calculus, and this is what we address in this paper.

With regards to applications, it is convenient to distinguish three
different notions of time-ordered exponentials when exponentiating quantum
stochastic integrals. The usual Dyson form, with It\^{o} differentials, does
not exponentiate a derivation-valued process to a homomorphic map. Instead
we must either use Stratonovich differentials or exponentiate over time
steps. The former strategy was applied using quantum white noises in a
series of papers by the author \cite{G}, \cite{G97}, while the latter was
investigated in \cite{HP1} and \cite{H}. These alternatives to the
It\^{o}-Dyson form coincide in the special case of quantum diffusions.

\bigskip

Let $\frak{A}$ be the *-algebra of operators modelling a system and its
environment. The space of all linear maps $L:$by $\frak{A}\mapsto \frak{A}$,
having the reality property $L\left( X^{\dagger }\right) =L\left( X\right)
^{\dagger }$, will be denoted by $\mathcal{L}\left( \frak{A}\right) $. The 
\textit{dissipation} of such a map is defined to be the bilinear mapping $%
\frak{D}_{L}:\frak{A}\times \frak{A}\mapsto \frak{A}$ given by 
\begin{equation}
\frak{D}_{L}\left( X,Y\right) =L(XY)-\left( LX\right) Y-X\left( LY\right) .
\label{a:dissipation}
\end{equation}
If $\frak{D}_{L}$ is zero then $L$ is called a \textit{derivation}. If $v\in 
\mathcal{L}\left( \frak{A}\right) $ is a derivation on the algebra if we
have the Leibniz property 
\begin{equation*}
v\left( XY\right) =v\left( X\right) Y+Xv\left( Y\right) ,
\end{equation*}
for all $X,Y\in \frak{A}$. Let us introduce the notations 
\begin{equation*}
v^{\circ n}=\underset{n\text{ times}}{\underbrace{v\circ \cdots \circ v}}%
,\quad e^{tv}\left( \cdot \right) =\sum_{n=0}^{\infty }\frac{t^{n}}{n!}%
v^{\circ n}\left( \cdot \right) ,
\end{equation*}
then $v^{\circ n}\left( XY\right) =\sum_{m}\binom{n}{m}v^{\circ n}\left(
X\right) v^{\circ n-m}\left( Y\right) $ and 
\begin{equation}
e^{tv}\left( XY\right) =e^{tv}\left( X\right) Y+Xe^{tv}\left( Y\right) .
\end{equation}
Therefore, derivations act as the generators of flow maps that preserve the
algebraic structure (homomorphisms).

We introduce the following causal structure: for each $t\geq 0$, let $\frak{A%
}_{t]}$ be the *-subalgebra of $\frak{A}$ modelling the system and its
environment up to time $t$, then we assume that we have the isotony
condition $\frak{A}_{s]}\subset \frak{A}_{t]}$ whenever $s<t$. The family $%
\left\{ \frak{A}_{t]}:t\geq 0\right\} $ is called a \textit{filtration}. Let 
$\mathbb{E}_{t]}:\frak{A}\mapsto \frak{A}_{t]}$ be a projective conditional
expectation

A flow is a family $\left\{ \Phi _{t,s}:t\geq s\geq 0\right\} $ of maps in $%
\mathcal{L}\left( \frak{A}\right) $ with the properties

\begin{itemize}
\item[i)]  $\Phi _{t,r}\circ \Phi _{r,s}=\Phi _{t,s}$, whenever $t\geq r\geq
s;$

\item[ii)]  $\lim_{t\downarrow s}\Phi _{t,s}=$ $id$, the identity map on $%
\frak{A}$.
\end{itemize}

The flow is said to be \textit{adapted }to the filtration if $\Phi
_{t,s}\left( \frak{A}_{s]}\right) \subset \frak{A}_{t]}$ whenever $s<t$ and
is said to be \textit{homomorphic} if $\Phi _{t,s}\left( XY\right) =\Phi
_{t,s}\left( X\right) \Phi _{t,s}\left( Y\right) $ for all $X,Y\in \frak{A}%
_{s]}$.

Given a flow, we define the maps $\left\{ L_{t,s}:t\geq s\geq 0\right\} $ of
maps in $\mathcal{L}\left( \frak{A}\right) $ by $L_{t,s}=\Phi _{t,s}-id$.
Let us write $dL_{t}$ for $L_{t+dt,t}$ and we may think of this as an $%
\mathcal{L}\left( \frak{A}\right) $-valued measure which we refer to as a 
\textit{stochastic generator}. Given two such measures $dL_{t}$ and $%
dL_{t}^{\prime }$, we define their mutual quadratic variation over time
interval $\left[ S,T\right] $ as 
\begin{equation*}
\int_{S}^{T}dL_{t}\circ dL_{t}^{\prime }=\lim_{|P\left( S,T\right)
|\rightarrow 0^{+}}\sum_{\left( t_{j},t_{j+1}\right) \in P\left( S,T\right)
}L\left( t_{j+1}+t_{j},t_{j}\right) \circ L^{\prime }\left(
t_{j+1}+t_{j},t_{j}\right)
\end{equation*}
where we have the limit over all partitions $P\left( S,T\right) $ of $\left[
S,T\right] $ into sub-intervals of maximum length $|P\left( S,T\right) |$.
In general, for a family $\left\{ K_{t,s}:t\geq s\geq 0\right\} $ of maps in 
$\mathcal{L}\left( \frak{A}\right) $, if we have that, for every finite
interval $\left[ S,T\right] $, 
\begin{equation*}
\lim_{|P\left( S,T\right) |\rightarrow 0^{+}}\sum_{\left(
t_{j},t_{j+1}\right) \in P\left( S,T\right) }K\left(
t_{j+1}+t_{j},t_{j}\right) =0
\end{equation*}
then we write $dK_{r}=o\left( dt\right) $. We say that a flow is regular, or
deterministic, if $dL_{t}\circ dL_{t}=o\left( dt\right) $. In general,
however, this does not hold and we typically encounter the It\^{o} rule for
differentials of compositions, viz. 
\begin{equation*}
d\left( L\circ L^{\prime }\right) \equiv \left( dL\right) \circ L^{\prime
}+L\circ \left( dL^{\prime }\right) +\left( dL\right) \circ \left(
dL^{\prime }\right)
\end{equation*}
with the last term, the It\^{o} correction, being non-zero. (Here ``$\equiv $%
'' means equal up to terms of order $o\left( dt\right) $.)

A flow $\Phi $\ is said to admit a forward derivative $u_{t}\left( \cdot
\right) \in \mathcal{L}\left( \frak{A}\right) $ if the following limits
exist 
\begin{equation}
u_{t}\left( X\right) :=\lim_{\tau \rightarrow 0^{+}}\frac{1}{\tau }\mathbb{E}%
_{t]}\left[ \Phi _{t+\tau ,t}X-X\right]
\end{equation}
for each $t\geq 0,X\in \frak{A}_{t]}$. The forward derivative will generally
not be a derivation! If the flow admits a forward derivative $u_{t}$ then a
difference martingale $\mathcal{M}_{t,t_{0}}\left( \cdot \right) \in 
\mathcal{L}\left( \frak{A}_{t_{0}]}\right) $ is defined by 
\begin{equation*}
\Phi _{t,t_{0}}\left( X\right) =X+\int_{t_{0}}^{t}u_{s}\left( X\right) \,ds+%
\mathcal{M}_{t,t_{0}}\left( X\right) ,\qquad \left( t_{0}<t\right) .
\end{equation*}
For $s<r<t$, $\mathcal{M}_{t,s}=\mathcal{M}_{t,r}+\mathcal{M}_{r,s}$.

\subsection{It\^{o}-Dyson Exponentials}

The flow can be reconstructed from $dL_{t}$. We have the (It\^{o})
differential equations 
\begin{equation}
d\Phi _{t,t_{0}}\left( \cdot \right) =dL_{t}\circ \Phi _{t,t_{0}}\left(
\cdot \right)
\end{equation}
with initial condition $\lim_{t\downarrow t_{0}}\Phi _{t,t_{0}}=id$. We
write 
\begin{equation}
\Phi _{t,t_{0}}=\mathbf{\vec{T}}_{ID}\exp \left\{ \int_{t_{0}}^{t}dL\right\}
\end{equation}
where $\mathbf{\vec{T}}_{ID}$ is the Dyson time-ordering symbol and we refer
to right hand sides as \textit{It\^{o}-Dyson time-ordered exponentials}.
They can be developed as an, at least formal, Picard series 
\begin{eqnarray}
\Phi _{t,t_{0}}\left( X\right) &=&X+\int_{t_{0}}^{t}dL_{t_{1}}\left(
X\right) +\int_{t_{0}}^{t}dL_{t_{2}}\left(
\int_{t_{0}}^{t_{2}}dL_{t_{1}}\left( X\right) \right) +\cdots  \notag \\
&=&\sum_{n=0}^{\infty }\int_{\Delta _{n}\left( t_{0},t\right)
}dL_{t_{n}}\circ \cdots \circ dL_{t_{1}}\left( X\right) .
\end{eqnarray}
Here $\Delta _{n}\left( a,b\right) $ denotes the simplex consisting of all $%
n-$tuples $\left( t_{n},\cdots ,t_{1}\right) $ ordered so that $b\geq
t_{n}\geq \cdots \geq t_{1}\geq a$. So we have, again formally, 
\begin{equation*}
\mathbf{\vec{T}}_{ID}\exp \left\{ \int_{t_{0}}^{t}dL\right\}
=\sum_{n=0}^{\infty }\int_{\Delta _{n}\left( t_{0},t\right) }dL_{t_{n}}\circ
\cdots \circ dL_{t_{1}}.
\end{equation*}

Because the It\^{o} rule means a breakdown of the Leibniz identity, it
typically means that if $dL_{t}\left( \cdot \right) $ is equivalent to a
derivation, but is not regular, then the flow $\Phi $ we construct will fail
to be homomorphic. Indeed, the condition on $dL_{t}$ in order for the flow
to be a family homomorphisms is that its dissipation balances its
fluctuations \cite{AH}, viz. 
\begin{equation}
\frak{D}_{dL_{t}}=\left( dL_{t}\right) \otimes \left( dL_{t}\right) .
\end{equation}
More explicitly, $dL_{t}\left( XY\right)
=dL_{t}(X)Y+XdL_{t}(Y)+dL_{t}(X)dL_{t}(Y)$.

\subsection{Exponentiated Dyson Exponentials}

We may refer to $\Phi _{t,t_{0}}=e^{\left( t-t_{0}\right) v}$ as the
autonomous flow generated by velocity field $v$. More generally, we could
take a family $\left\{ v_{t}:t\geq 0\right\} $ of derivations and consider
the flow generated by $dL_{t}\left( \cdot \right) =v_{t}\left( \cdot \right)
dt$, that is, $v_{t}\left( \cdot \right) $ is the instantaneous velocity
field. This time, we can write the non-autonomous flow as $\Phi _{t,t_{0}}=%
\mathbf{\vec{T}}_{ID}\exp \left\{ \int_{t_{0}}^{t}v_{s}ds\right\} $. We
observe that 
\begin{equation*}
\Phi _{t+dt,t}\left( \cdot \right) =id\left( \cdot \right) +v_{t}\left(
\cdot \right) dt=e^{v\left( \cdot \right) dt}+o\left( dt\right) .
\end{equation*}

Unfortunately this cannot be the case when we consider differential
generators $dL_{t}$ that are not regular. The reason is that, because of the
It\^{o} calculus, compositions $\left( dL_{t}\right) ^{\circ n}$ need not be 
$o\left( dt\right) $ for $n\geq 2$. Let us write in general 
\begin{equation}
dH_{t}=e^{dL_{t}}-id=\sum_{n\geq 1}\left( dL_{t}\right) ^{\circ n}
\end{equation}
and define the\textit{\ exponentiated} \textit{Dyson}, or $ED$-type,
time-ordered exponential to be $\mathbf{\vec{T}}_{ED}\exp \left\{
\int_{t_{0}}^{t}dL\right\} =\mathbf{\vec{T}}_{ID}\exp \left\{
\int_{t_{0}}^{t}dH\right\} .$This time, if the $dL_{t}$ are derivations then
the $e^{dL_{t}}\left( \cdot \right) $ will behave as homomorphisms, and so
too will 
\begin{equation}
\mathbf{\vec{T}}_{ED}\exp \left\{ \int_{t_{0}}^{t}dL\right\} =\mathbf{\vec{T}%
}_{ID}\exp \left\{ \int_{t_{0}}^{t}\left( e^{dL}-id\right) \right\} .
\end{equation}

\subsection{Stratonovich-Dyson Exponentials}

As we have seen above, the It\^{o} calculus implies that the Dyson
time-ordered exponential $\mathbf{\vec{T}}_{ID}\exp \left\{
\int_{t_{0}}^{t}dL\right\} $ will not generally yield a homomorphism if the $%
dL_{t}\left( \cdot \right) $ are derivations. There is another strategy for
producing homomorphisms and that is to use the Stratonovich calculus instead
of the It\^{o} one. The It\^{o} rule (for ordinary products) is that $%
d\left( X_{t}Y_{t}\right) =X_{t}\left( dY_{t}\right) +\left( dX_{t}\right)
Y_{t}+\left( dX_{t}\right) \left( dY_{t}\right) $. The Leibniz rule may be
formally restored as $d\left( X_{t}Y_{t}\right) =X_{t}\ast \left(
dY_{t}\right) +\left( dX_{t}\right) \ast Y_{t}$ where we introduce
Stratonovich differentials $X_{t}\ast \left( dY_{t}\right) =X_{t}\left(
dY_{t}\right) +\frac{1}{2}\left( dX_{t}\right) \left( dY_{t}\right) $.

For compositions of infinitesimal generators we have the similar relation 
\begin{equation*}
d\left( X_{t}\circ Y_{t}\right) =X_{t}\circ \left( dY_{t}\right) +\left(
dX_{t}\right) \circ Y_{t}+\left( dX_{t}\right) \circ \left( dY_{t}\right) ,
\end{equation*}
with the Leibniz rule salvaged as 
\begin{equation*}
d\left( X_{t}\circ Y_{t}\right) =X_{t}\circledast \left( dY_{t}\right)
+\left( dX_{t}\right) \circledast Y_{t}
\end{equation*}
where now $X_{t}\circledast \left( dY_{t}\right) =X_{t}\circ \left(
dY_{t}\right) +\frac{1}{2}\left( dX_{t}\right) \circ \left( dY_{t}\right) $
and $\left( dX_{t}\right) \circledast Y_{t}=\left( dX_{t}\right) \circ Y_{t}+%
\frac{1}{2}\left( dX_{t}\right) \circ \left( dY_{t}\right) $.

We define the \textit{Stratonovich-Dyson}, or $SD$-type, time-ordered
exponential $\Phi _{t,t_{0}}=\mathbf{\vec{T}}_{SD}\exp \left\{
\int_{t_{0}}^{t}dL_{s}\right\} $ to be the solution to the
integro-differential equation 
\begin{equation*}
\Phi _{t,t_{0}}\left( X\right) =X+\int_{t_{0}}^{t}dG_{s}\,\left( \Phi
_{s,t_{0}}\left( X\right) \right)
\end{equation*}
where $dG_{t}$ is the Stratonovich differential 
\begin{equation*}
dG_{t}\circ \Phi =dL_{t}\circledast \Phi =dL_{t}\circ \Phi +\frac{1}{2}%
dL_{t}\circ dL_{t}\circ \Phi .
\end{equation*}
that is 
\begin{equation}
dG_{t}=dL_{t}+\frac{1}{2}\left( dL_{t}\right) ^{\circ 2}.
\end{equation}
We have now defined $\Phi _{t,t_{0}}=\mathbf{\vec{T}}_{SD}\exp \left\{
\int_{t_{0}}^{t}dL_{s}\right\} $ as the solution to 
\begin{equation*}
\Phi _{t+dt,t_{0}}\left( \cdot \right) =dL_{t}\circledast \Phi
_{t,t_{0}}\left( \cdot \right) =dG_{t}\circ \Phi _{t,t_{0}}\left( \cdot
\right)
\end{equation*}
with $\lim_{t\downarrow t_{o}}\Phi _{t,t_{0}}=id$. As the Leibniz rule is
observed in the Stratonovich calculus, $\mathbf{\vec{T}}_{SD}\exp \left\{
\int_{t_{0}}^{t}dL_{s}\right\} $ will be a homomorphism when the $dL_{t}$
are derivations.

\subsection{Properties}

If the flow is regular, say $dL_{t}\left( \cdot \right) \equiv v_{t}\left(
\cdot \right) dt$ for some $v_{t}\in \mathcal{L}\left( \frak{A}\right) $ and 
$\left( dL_{t}\right) ^{\circ n}\equiv 0$ for $n\geq 2$, then the various
notion of time-ordered exponentials coincide. We say that the flow is a 
\textit{diffusion} if $\left( dL_{t}\right) ^{\circ 2}$ is equivalent to $%
A_{t}\left( \cdot \right) dt$ for some $A_{t}\in \mathcal{L}\left( \frak{A}%
\right) $, however, $\left( dL_{t}\right) ^{\circ n}\equiv 0$ for $n\geq 3$.
In this case, we have the truncation 
\begin{equation*}
dH_{t}=e^{dL_{t}}-id=dL_{t}+\frac{1}{2}\left( dL_{t}\right) ^{\circ 2}
\end{equation*}
and so $dH_{t}=dG_{t}$ and therefore 
\begin{equation*}
\mathbf{\vec{T}}_{ED}\exp \left\{ \int_{t_{0}}^{t}dL_{s}\right\} =\mathbf{%
\vec{T}}_{SD}\exp \left\{ \int_{t_{0}}^{t}dL_{s}\right\} .
\end{equation*}
This is however a fluke which cannot be expected to hold for stochastic
processes other than diffusions.

\bigskip

The $ED$-type exponential occurs when we consider discrete
time-approximations. We have for instance 
\begin{eqnarray*}
\mathbf{\vec{T}}_{ED}\exp \left\{ \int_{t_{0}}^{t}dL\right\} &=&\lim_{\max
\left| t_{j+1}-t_{j}\right| \rightarrow 0}\exp \left\{ \Phi
_{t_{N},t_{N-1}}\right\} \circ \cdots \circ \exp \left\{ \Phi
_{t_{1},t_{0}}\right\} \\
&=&\lim_{\max \left| t_{j+1}-t_{j}\right| \rightarrow 0}\exp \left\{
\int_{t_{N-1}}^{t_{N}}dL_{s}\right\} \circ \cdots \circ \exp \left\{
\int_{t_{0}}^{t_{1}}dL_{s}\right\}
\end{eqnarray*}
where the limit is over all partitions $t=t_{N}>t_{N-1}>\cdots >t_{1}>t_{0}$%
. Let us suppose that the stochastic generator has essentially commutative
increments, that is, 
\begin{equation*}
\left[ \Phi _{t,s},\Phi _{t^{\prime },s^{\prime }}\right] =0
\end{equation*}
whenever the intervals $\left( s,t\right) $ and $\left( s^{\prime
},t^{\prime }\right) $ do not overlap, then 
\begin{equation*}
\mathbf{\vec{T}}_{ED}\exp \left\{ \int_{t_{0}}^{t}dL\right\} =\exp \Phi
_{t,t_{0}}.
\end{equation*}

\bigskip

The $SD$-type exponential occurs when we wish to approximate stochastic
flows by regular ones. Let $v_{t}^{\left( \lambda \right) }\left( \cdot
\right) $ be velocity fields parametrized by $t\geq 0$ and $\lambda >0$.
Suppose that we have the limit 
\begin{equation*}
\lim_{\lambda \rightarrow 0}\int_{t_{0}}^{t}v_{s}^{\left( \lambda \right)
}\left( \cdot \right) ds=\int_{t_{0}}^{t}dL_{s}\left( \cdot \right)
\end{equation*}
where $dL_{t}$ is possibly stochastic. We may then expect the limit 
\begin{equation*}
\lim_{\lambda \rightarrow 0}\mathbf{\vec{T}}_{ID}\exp \left\{
\int_{t_{0}}^{t}v_{s}^{\left( \lambda \right) }ds\right\} =\mathbf{\vec{T}}%
_{SD}\exp \left\{ \int_{t_{0}}^{t}dL_{s}\right\} .
\end{equation*}
Evidently we should not have expected the Dyson exponential on the right
hand side as limit should be a homomorphism. In the theory of approximating
stochastic differential equations by ordinary differential equations, it is
well known that the Stratonovich calculus is the one that best anticipates
the limit form.

\subsection{Examples from Mathematical Physics.}

\paragraph{Classical mechanics}

Let $\frak{A}$ be the C$^{\infty }$ functions on a manifold $M$. Here $\frak{%
A}$ is a commutative algebra (though not generally a C*-algebra) with
respect to pointwise multiplication, and derivations correspond to the
tangent vector fields. In this case, the dissipation is known by several
different names: the Gamma operator, l'op\'{e}rateur carr\'{e} du champ, the
cometric operator, etc. (see Meyer's appendix to \cite{Emery}).

If, however, $M$ is also endowed with a Poisson brackets which is, of
course, a bilinear mapping. We may take our product to be the
anti-symmetric, non-associative one given by the choice $f\star g\equiv
\left\{ f,g\right\} $. In this case the $\star $-automorphisms are those
maps that preserve the Poisson brackets. Likewise $\frak{D}_{\star }L$ will
now determine the extent to which a semi-group $e^{tL}$ destroys the Poisson
structure: explicitly, we have $\frak{D}_{L}\left( f,g\right) =L\left(
\left\{ f,g\right\} \right) -\left\{ Lf,g\right\} -\left\{ f,Lg\right\} $.
Given a real function $h\in \frak{A}$, the Hamiltonian vector field $X_{h}$
generated by $h$ is defined by $X_{h}\left( f\right) :=\left\{ f,h\right\} $%
. By the Jacobi property of Poisson brackets, $X_{h}$ will be a
Poisson-derivation 
\begin{equation*}
X_{h}\left\{ f,g\right\} =\left\{ X_{h}f,g\right\} +\left\{ f,X_{h}g\right\}
;
\end{equation*}
and by the Leibniz property of Poisson brackets it will also be a tangent
vector field. The Poisson manifold is said to be Poisson-simple if the only
maps having Poisson-dissipation zero are the Hamiltonian vector fields \cite
{Weinstein}. For symplectic manifolds, all Poisson derivations are locally
Hamiltonian generated.

\paragraph{Quantum Mechanics}

Next let $\frak{A}$ be an algebra of operators acting on a Hilbert space $%
\mathcal{H}$, with the taking of adjoints $\dagger $ as the usual
involution. The operator product is non-commutative and to a certain extent
carries out the role played by both pointwise multiplication and Poisson
brackets in classical mechanics. The linear maps on $\frak{A}\mapsto \frak{A}
$ are sometimes referred to as super-operators. For a self-adjoint operator $%
h\in \frak{A}$, we define the super-operator $X_{h}$by $X_{h}\left( f\right)
:=\frac{1}{i}\left[ f,h\right] $. If $\frak{A}$ is a W$^{\ast }$-algebra, or
a unital simple C$^{\ast }$- algebra, then it is well-known that all real
derivations take this Hamiltonian form \cite{Sakai}. The dissipation of a
super-operator, with respect to the operator product, was introduced by
Lindblad \cite{Lindblad} as a key ingredient in analyzing generators of
completely positive semi-groups.

\bigskip

\section{Quantum Stochastic Flows}

Let $\frak{H}=\Gamma \left( L^{2}\left( \mathbb{R}^{+},dt\right) \right) $
be the Fock space over square-integrable functions of positive time and let $%
\frak{H}_{t}=\Gamma \left( L^{2}\left( \left[ 0,t\right] ,dt\right) \right) $%
. The family $\left\{ \frak{H}_{t}:t\geq 0\right\} $ then gives a filtration
of Hilbert subspaces of $\frak{H}$. Fixing an initial Hilbert space $\frak{h}
$, we consider the filtration of $\frak{A=B}\left( \frak{h}\otimes \frak{H}%
\right) $ specified by 
\begin{equation*}
\frak{A}_{t]}=\frak{B}\left( \frak{h}\otimes \frak{H}_{t]}\right) .
\end{equation*}
The quartet of fundamental quantum stochastic processes \cite{HP}\ may be
denoted by $\left\{ A_{t}^{\alpha \beta }:t\geq 0\right\} $: these are $%
A_{t}^{00}=t$ (time), $A_{t}^{10}=A_{t}$ (creation), $A_{t}^{10}=A_{t}^{\dag
}$ (annihilation) and $A_{t}^{11}=\Lambda _{t}$ (conservation). Their
It\^{o} table is then 
\begin{equation*}
dA_{t}^{\alpha 1}dA_{t}^{1\beta }=dA_{t}^{\alpha \beta },
\end{equation*}
with all other second order differentials vanishing. This may be written
explicitly as \cite{HP} 
\begin{equation*}
\begin{tabular}{l|llll}
$\times $ & $dA^{\dag }$ & $d\Lambda $ & $dA$ & $dt$ \\ \hline
$dA^{\dag }$ & $0$ & $0$ & $0$ & $0$ \\ 
$d\Lambda $ & $dA^{\dag }$ & $d\Lambda $ & $0$ & $0$ \\ 
$dA$ & $dt$ & $dA$ & $0$ & $0$ \\ 
$dt$ & $0$ & $0$ & $0$ & $0$%
\end{tabular}
\end{equation*}
We have the adjoint relations $\left( A_{t}^{\alpha \beta }\right) ^{\dag
}=A_{t}^{\beta \alpha }$: that is, $\Lambda _{t}$ is self-adjoint and $%
A_{t}^{\dag }$ is indeed the adjoint of $A_{t}$ We remark that the
combinations $Q_{t}=A_{t}+A_{t}^{\dag }$ and $N_{t}=\Lambda
_{t}+A_{t}+A_{t}^{\dag }+t$ give representations for the Wiener and Poisson
processes respectively when we specify the Fock vacuum as state.

\bigskip

A closed evolution restricted to $\frak{A}_{0]}=\frak{B}\left( \frak{h}%
\right) $ may be described by the family of unitaries 
\begin{equation*}
U_{t,t_{0}}=\mathbf{\vec{T}}_{ID}\exp \left\{
-i\int_{t_{0}}^{t}H_{s}ds\right\}
\end{equation*}
where $\left\{ H_{s}:s\geq 0\right\} $ is a family of self-adjoint operators
forming what we usually call a time-dependent Hamiltonian.

Our aim is to construct unitary quantum stochastic processes on $\frak{A}$
of the form 
\begin{equation}
U_{t,t_{0}}=\mathbf{\vec{T}}_{ID}\exp \left\{
-i\int_{t_{0}}^{t}dG_{s}\right\} =\mathbf{\vec{T}}_{SD}\exp \left\{
-i\int_{t_{0}}^{t}dE_{s}\right\}
\end{equation}
where 
\begin{eqnarray*}
dG_{t} &=&\mathsf{G}_{\alpha \beta }\otimes dA_{t}^{\alpha \beta }=\mathsf{G}%
_{00}\otimes dt+\mathsf{G}_{10}\otimes dA_{t}^{\dag }+\mathsf{G}_{01}\otimes
dA_{t}+\mathsf{G}_{11}\otimes d\Lambda _{t} \\
dE_{t} &=&\mathsf{E}_{\alpha \beta }\otimes dA_{t}^{\alpha \beta }=\mathsf{E}%
_{00}\otimes dt+\mathsf{E}_{10}\otimes dA_{t}^{\dag }+\mathsf{E}_{01}\otimes
dA_{t}+\mathsf{E}_{11}\otimes d\Lambda _{t}.
\end{eqnarray*}
(We use a summation convention that repeated Greek indices are summed over
values $0$ and $1$.) We take the coefficients $\mathsf{G}_{\alpha \beta }$
and $\mathsf{E}_{\alpha \beta }$ to be bounded operators on $\frak{h}$.
Special choices of the coefficients should lead to stochastic evolutions
driving by either Wiener or Poisson Noise, however, it is known that
classical stochastic processes do not account for all the quantum stochastic
evolutions we would wish to consider, and so we work with all four
fundamental processes.

\bigskip

\subsection{Quantum Stratonovich Calculus}

We do not aim a full generalization of the Stratonovich prescription to
quantum stochastic calculus, however, we shall give the algebraic rules
which tell us how to transform certain It\^{o} integrals into Stratonovich
ones in a manner that parallels the classical theory. There are two main
surprises: the first being that it can be done; the second being that there
is an ambiguity in the definition.

Let $\left\{ X_{t}:t\geq 0\right\} $ be a family of operators (a quantum
stochastic process) on $\frak{h}\otimes \frak{H}$. If we have that $X_{t}$
acts nontrivially only on the subspace$\frak{h}\otimes \frak{H}_{t]}$ then
we say that the process is adapted. Our definition of Stratonovich
differentials will amount to 
\begin{eqnarray}
\left( dX_{t}\right) \ast X_{t} &=&\left( dX_{t}\right) Y_{t}+\kappa \left(
dX_{t}\right) \left( dY_{t}\right) ,  \notag \\
X_{t}\ast \left( dY_{t}\right) &=&X_{t}\left( dY_{t}\right) +\kappa ^{\ast
}\left( dX_{t}\right) \left( dY_{t}\right) ,
\end{eqnarray}
where $\kappa $ is a complex number with $\func{Re}\kappa =\dfrac{1}{2}$.
Evidently we have $\left( dX_{t}\right) \ast X_{t}+X_{t}\ast \left(
dY_{t}\right) =d\left( X_{t}Y_{t}\right) $ from the quantum It\^{o} formula 
\cite{HP}. The fact that we may choose the imaginary part of $\kappa $ means
that we have a ``gauge freedom'' which in many respects is similar to that
in the Tomita-Takesaki theory. The symmetric choice would be $\kappa =\dfrac{%
1}{2}$, however, its explanation is as a damping constant, and physically
this may be complex.

The stochastic Schr\"{o}dinger equation\ from above is then (ignoring the $%
t_{0}$ dependence) 
\begin{equation}
dU_{t}=-i\left( dG_{t}\right) U_{t}=-i\left( dE_{t}\right) \ast U_{t}
\end{equation}
and we would like to determine how to transform between It\^{o} and
Stratonovich forms. Let us begin by computing $\left( dA_{t}^{\alpha \beta
}\right) \ast U_{t}=\left( dA_{t}^{\alpha \beta }\right) U_{t}+\kappa \left(
dA_{t}^{\alpha \beta }\right) \left( dU_{t}\right) $. We have 
\begin{eqnarray*}
\left( dA_{t}^{\alpha \beta }\right) \left( dU_{t}\right) &=&-i\left(
dA_{t}^{\alpha \beta }\right) \left( \mathsf{G}_{\mu \nu }\otimes
dA_{t}^{\mu \nu }\right) U_{t} \\
&=&-i\delta _{1\beta }\left( \mathsf{G}_{1\nu }\otimes dA_{t}^{\alpha \nu
}\right) U_{t}
\end{eqnarray*}
from which we see that 
\begin{equation*}
\left\{ 
\begin{array}{l}
\left( d\Lambda _{t}\right) \ast U_{t}=\left( 1-i\kappa \mathsf{G}%
_{11}\right) \left( d\Lambda _{t}\right) U_{t}-i\kappa \mathsf{G}_{10}\left(
dA_{t}^{\dag }\right) U_{t} \\ 
\left( dA_{t}^{\dag }\right) \ast U_{t}=\left( dA^{\dag }\right) U_{t} \\ 
\left( dA_{t}\right) \ast U_{t}=\left( 1-i\kappa \mathsf{G}_{11}\right)
\left( dA_{t}\right) U_{t}-i\kappa \mathsf{G}_{10}\left( dt\right) U_{t} \\ 
\left( dt\right) \ast U_{t}=\left( dt\right) U_{t}
\end{array}
\right.
\end{equation*}
and inversely 
\begin{equation*}
\left\{ 
\begin{array}{l}
\left( dt\right) U_{t}=\left( dt\right) \ast U_{t} \\ 
\left( dA^{\dag }\right) U_{t}=\left( dA_{t}^{\dag }\right) \ast U_{t} \\ 
\left( dA_{t}\right) U_{t}=\left( 1-i\kappa \mathsf{G}_{11}\right) ^{-1}%
\left[ \left( dA_{t}\right) \ast U_{t}+i\kappa \mathsf{G}_{10}\left(
dt\right) U_{t}\right] \\ 
\left( d\Lambda _{t}\right) U_{t}=\left( 1-i\kappa \mathsf{G}_{11}\right)
^{-1}\left[ \left( d\Lambda _{t}\right) \ast U_{t}+i\kappa \mathsf{G}%
_{10}\left( dA_{t}^{\dag }\right) U_{t}\right]
\end{array}
\right. .
\end{equation*}
We can next of all read off the relationship between the $\mathsf{G}_{\alpha
\beta }$ and the $\mathsf{E}_{\alpha \beta }$ by comparing coefficients: 
\begin{equation*}
\left\{ 
\begin{array}{l}
\mathsf{E}_{11}=\dfrac{\mathsf{G}_{11}}{1-i\kappa \mathsf{G}_{11}} \\ 
\mathsf{E}_{10}=\dfrac{1}{1-i\kappa \mathsf{G}_{11}}\mathsf{G}_{10} \\ 
\mathsf{E}_{01}=\mathsf{G}_{01}\dfrac{1}{1-i\kappa \mathsf{G}_{11}} \\ 
\mathsf{E}_{00}=\mathsf{G}_{00}+i\kappa \mathsf{G}_{01}\dfrac{1}{1-i\kappa 
\mathsf{G}_{11}}\mathsf{G}_{10}
\end{array}
\right.
\end{equation*}
or inversely 
\begin{equation*}
\left\{ 
\begin{array}{l}
\mathsf{G}_{11}=\dfrac{\mathsf{E}_{11}}{1+i\kappa \mathsf{E}_{11}} \\ 
\mathsf{G}_{10}=\dfrac{1}{1+i\kappa \mathsf{E}_{11}}\mathsf{E}_{10} \\ 
\mathsf{G}_{01}=\mathsf{E}_{01}\dfrac{1}{1+i\kappa \mathsf{E}_{11}} \\ 
\mathsf{G}_{00}=\mathsf{E}_{00}+i\kappa \mathsf{E}_{01}\dfrac{1}{1+i\kappa 
\mathsf{E}_{11}}\mathsf{E}_{10}
\end{array}
\right. .
\end{equation*}
The relationship between the It\^{o} and Stratonovich coefficients may be
written more compactly using the following remarkable formulas 
\begin{eqnarray}
\mathsf{E}_{\alpha \beta } &=&\mathsf{G}_{\alpha \beta }+i\kappa \mathsf{G}%
_{\alpha 1}\dfrac{1}{1-i\kappa \mathsf{G}_{11}}\mathsf{G}_{1\beta },  \notag
\\
\mathsf{G}_{\alpha \beta } &=&\mathsf{E}_{\alpha \beta }-i\kappa \mathsf{E}%
_{\alpha 1}\dfrac{1}{1+i\kappa \mathsf{E}_{11}}\mathsf{E}_{1\beta },
\end{eqnarray}
showing a duality between them. In particular the operators $1-i\kappa 
\mathsf{G}_{11}$ and $1+i\kappa \mathsf{E}_{11}$ are required to be
invertible and inverse to each other.

\bigskip

The conditions for unitarity are the isometric and co-isometric properties $%
U_{t}^{\dag }U_{t}=1=U_{t}U_{t}^{\dag }$. In differential terms this is 
\begin{eqnarray}
0 &=&d\left( U_{t}^{\dag }U_{t}\right) =\left( dU_{t}^{\dag }\right)
U_{t}+U_{t}^{\dag }\left( dU_{t}\right) +\left( dU_{t}^{\dag }\right) \left(
dU_{t}\right)  \notag \\
&=&\left( dU_{t}^{\dag }\right) \ast U_{t}+U_{t}^{\dag }\ast \left(
dU_{t}\right)
\end{eqnarray}
with $0=d\left( U_{t}U_{t}^{\dag }\right) $ similarly. From the Stratonovich
form $dU_{t}=-i\left( dE_{t}\right) \ast U_{t}$ we see that it is enough to
ask that $\left( dE_{t}\right) ^{\dag }=dE_{t}$. The unitarity of the
process should come down to the conditions 
\begin{equation}
\left( \mathsf{E}_{\alpha \beta }\right) ^{\dag }=\mathsf{E}_{\beta \alpha }.
\end{equation}
The process $U_{t,t_{0}}=\mathbf{\vec{T}}_{SD}\exp \left\{
-i\int_{t_{0}}^{t}dE_{s}\right\} $ is evidently unitary because we are using
the Stratonovich calculus to time-order explicitly unitary components. If we
use the quantum It\^{o} calculus with $dU_{t}=-i\left( dG_{t}\right) U_{t}$\
we see that 
\begin{equation*}
0=d\left( U_{t}^{\dag }U_{t}\right) =U_{t}^{\dag }\left[ i\left(
dG_{t}^{\dag }\right) -i\left( dG_{t}\right) -\frac{1}{2}\left(
dG_{t}\right) ^{\dag }\left( dG_{t}\right) \right] U_{t}
\end{equation*}
which implies 
\begin{eqnarray*}
0 &=&i\left( dG_{t}^{\dag }\right) -i\left( dG_{t}\right) -\frac{1}{2}\left(
dG_{t}\right) ^{\dag }\left( dG_{t}\right) \\
&=&\left( i\mathsf{G}_{\beta \alpha }^{\dag }-i\mathsf{G}_{\alpha \beta }-%
\frac{1}{2}\mathsf{G}_{1\alpha }^{\dag }\mathsf{G}_{1\beta }\right) \otimes
dA_{t}^{\alpha \beta }.
\end{eqnarray*}
It turns out that the four equations $0=i\mathsf{G}_{\beta \alpha }^{\dag }-i%
\mathsf{G}_{\alpha \beta }-\frac{1}{2}\mathsf{G}_{1\alpha }^{\dag }\mathsf{G}%
_{1\beta }$ guarantee both the isometric and co-isometric properties, and
therefore unitarity. It is well-known that the general solution to this
equation is \cite{HP} 
\begin{eqnarray*}
\mathsf{G}_{00} &=&\mathsf{H}-i\frac{1}{2}\mathsf{K}^{\dag }\mathsf{K} \\
\mathsf{G}_{10} &=&\mathsf{K} \\
\mathsf{G}_{01} &=&\mathsf{K}^{\dagger }\mathsf{W} \\
\mathsf{G}_{11} &=&i\left( \mathsf{W}-1\right)
\end{eqnarray*}
with $\mathsf{W}$ unitary, $\mathsf{H}$ self-adjoint, and $\mathsf{K}$
bounded but otherwise arbitrary. One can readily check that the coefficients 
$\mathsf{G}_{\alpha \beta }$ will satisfy these conditions once $\left( 
\mathsf{E}_{\alpha \beta }\right) ^{\dag }=\mathsf{E}_{\beta \alpha }$ with
the explicit choices 
\begin{eqnarray*}
\mathsf{W} &=&\frac{1-i\kappa ^{\ast }\mathsf{E}_{11}}{1+i\kappa \mathsf{E}%
_{11}}, \\
\mathsf{K} &=&\frac{1}{1+i\kappa \mathsf{E}_{11}}\mathsf{E}_{10}, \\
\mathsf{H} &=&\mathsf{E}_{00}+\func{Im}\kappa \mathsf{E}_{01}\frac{1}{%
1+i\kappa \mathsf{E}_{11}}\mathsf{E}_{10}.
\end{eqnarray*}
So, although the stochastic generator $dG_{t}$ does not look self-adjoint,
the process $U_{t,t_{0}}=\mathbf{\vec{T}}_{ID}\exp \left\{
-i\int_{t_{0}}^{t}dG_{s}\right\} $ is nevertheless unitary.

\bigskip

The flow is then given by the family of maps 
\begin{equation*}
\Phi _{t,t_{0}}\left( X\right) =U_{t,t_{0}}^{\dag }\left( X\right)
U_{t,t_{0}}
\end{equation*}
which defines a system of homomorphic flows on $\frak{A}$. The quantum
stochastic differential equation for $\Phi _{t,t_{0}}\left( X\right) $ is 
\begin{equation*}
d\Phi _{t,t_{0}}=dL_{t}\circ \Phi _{t,t_{0}}
\end{equation*}
with 
\begin{equation*}
dL_{t}\left( \cdot \right) =\mathcal{L}_{\alpha \beta }^{t}\left( \cdot
\right) \otimes dA_{t}^{\alpha \beta }
\end{equation*}
where $\mathcal{L}_{\alpha \beta }^{t}\left( \mathsf{X}\right) =i\left( 
\mathsf{G}_{\beta \alpha }^{t}\right) ^{\dag }\mathsf{X}-i\mathsf{X}\left( 
\mathsf{G}_{\alpha \beta }^{t}\right) -\frac{1}{2}\left( \mathsf{G}_{1\alpha
}^{t}\right) ^{t}\mathsf{X}\left( \mathsf{G}_{1\beta }^{t}\right) $ and $%
\mathsf{G}_{\alpha \beta }^{t}=\Phi _{t,t_{0}}\left( \mathsf{G}_{\alpha
\beta }\right) $. The stochastic derivation property is then stated in the
form 
\begin{equation*}
\frak{D}\mathcal{L}_{\alpha \beta }=\mathcal{L}_{\alpha 1}\otimes \mathcal{L}%
_{1\beta };
\end{equation*}
that is, $\mathcal{L}_{\alpha \beta }\left( XY\right) -\mathcal{L}_{\alpha
\beta }\left( X\right) Y-X\mathcal{L}_{\alpha \beta }\left( Y\right) =%
\mathcal{L}_{\alpha 1}\left( X\right) \mathcal{L}_{1\beta }\left( Y\right) .$
These are the well-known structure equations \ for non-commutative flows 
\cite{EVHUD}.

\subsection{Remarks}

In principle, the algebraic manipulations can be extended to time-dependent $%
\mathsf{E}_{\alpha \beta }\in \frak{B}\left( \frak{h}\right) $ and more
generally to adapted coefficients. We may also generalize to $N$ Bose noises 
$A_{t}^{\alpha \beta }$ where now the Greek indices run over $0,1,\cdots N$
and this leads to a tensorial version of the equations (14). In this case we
are free to introduce additional gauge degrees of freedom.

We mention that we have the following approximations theorem which is the
quantum analogue of the Wong-Zakai result and which justifies our
construction of quantum Stratonovich calculus. Let $a_{t}^{\sharp }\left(
\lambda \right) $ be Bose fields on a Fock space $\frak{H}_{R}^{\left(
\lambda \right) }$ correspond to some physical system which we shall refer
to as the reservoir. we have chosen to parameterize them by time $t$ and
also a scale parameter $\lambda >0$. For $\lambda $ fixed, we consider
canonical commutation relations of the type 
\begin{equation}
\left[ a_{t}\left( \lambda \right) ,a_{s}^{\dag }\left( \lambda \right) %
\right] =G_{\lambda }\left( t-s\right)
\end{equation}
where $G_{\lambda }$ is the two point function and is assumed to be a
regular function of the time difference. We shall assumed that $G_{\lambda
}\left( \cdot \right) $ is integrable with $\int_{-\infty }^{\infty
}G_{\lambda }=1$ and we naturally require that 
\begin{equation*}
G_{\lambda }\left( -\tau \right) =G_{\lambda }\left( \tau \right) ^{\ast }.
\end{equation*}
Let $\kappa =\int_{0}^{\infty }G_{\lambda }$ then $\func{Re}\kappa =\frac{1}{%
2}$ and $\int_{-\infty }^{0}G_{\lambda }=\kappa ^{\ast }$.

We now consider what happens if, in the limit $\lambda \rightarrow 0$, we
have 
\begin{equation}
G_{\lambda }\left( \tau \right) \rightarrow \delta \left( \tau \right)
\end{equation}
in the sense of Schwartz distribution. In particular, we wish to study the
asymptotic behavior of the unitary $U_{t}\left( \lambda \right) $ coupling a
given system with state space $\frak{h}$ to the reservoir with an
interaction Hamiltonian $\Upsilon _{t}\left( \lambda \right) $. Here $%
U_{t}\left( \lambda \right) $ is given as the solution to 
\begin{equation*}
U_{t}\left( \lambda \right) =1-i\int_{0}^{t}\Upsilon _{s}\left( \lambda
\right) U_{s}\left( \lambda \right) ds.
\end{equation*}

\begin{theorem}
\cite{GQCLT} \textit{For the interactions on }$\frak{h}\otimes \frak{H}%
_{R}^{\left( \lambda \right) }$\textit{\ of the type} 
\begin{equation}
\Upsilon _{t}\left( \lambda \right) =\mathsf{E}_{11}\otimes a_{t}^{\dag
}\left( \lambda \right) a_{t}\left( \lambda \right) +\mathsf{E}_{10}\otimes
a_{t}^{\dag }\left( \lambda \right) +\mathsf{E}_{01}\otimes a_{t}\left(
\lambda \right) +\mathsf{E}_{00}\otimes 1
\end{equation}
\textit{with }$\mathsf{E}_{\alpha \beta }\in \mathcal{B}\left( \frak{h}%
\right) $\textit{, }$\mathsf{E}_{11}$\textit{\ and }$\mathsf{E}_{00}$\textit{%
\ self-adjoint, }$\mathsf{E}_{10}=\mathsf{E}_{01}^{\dagger }$, \textit{and }$%
\left\| \kappa \mathsf{E}_{11}\right\| <1,$\textit{\ the weak matrix limit
of }$U_{t}\left( \lambda \right) $\textit{\ is described by the unitary
quantum stochastic process }$U_{t}$ \textit{given by the Stratonovich-Dyson
time-ordered exponential} 
\begin{equation*}
U_{t}=\mathbf{\vec{T}}_{SD}\exp \left\{ -i\int_{t_{0}}^{t}dE_{s}\right\}
\end{equation*}
with $dE_{t}=\mathsf{E}_{\alpha \beta }\otimes dA_{t}^{\alpha \beta }$.
\end{theorem}

\bigskip

The condition $\left\| \kappa \mathsf{E}_{11}\right\| <1$ is required to
ensure that multiple scatterings diminish rather than augment amplitudes -
it also allows (14) to be expanded in a geometric series. A simpler version
of this result, applicable for commuting coefficients only, was given in 
\cite{C}. The convergence also applies to the Heisenberg dynamics and so we
get convergence of the regular pre-limit flow to the stochastic quantum
flow. A similar set of formula arise when the $a_{t}^{\sharp }\left( \lambda
\right) $ are replaced by Fermion fields \cite{GS05a}: the fundamental
processes $A_{t}^{\alpha \beta }$ now being the Fermi analogues.

\subsection{Addition Rules for Stochastic Derivations}

Let $\left\{ dL_{t}^{\left( n\right) }\right\} _{n}$ be a finite collection
of stochastic derivations: their sum is not typically another stochastic
derivation. In general, $\sum_{n}dL_{t}^{\left( n\right) }+dF$ defines a
stochastic derivation only for some suitable choice of ``It\^{o}''
correction $F$. For the quantum problem, we realize each stochastic
derivation $dL_{t}$\ as a function of the operators $\mathsf{E}_{\alpha
\beta },$ i.e. $dL_{t}=dG_{t}\left( \mathsf{E}_{\alpha \beta }\right) $.

The natural procedure is then to consider the total stochastic Hamiltonian $%
dE_{t}=\sum_{n}dE_{t}^{\left( n\right) }=\sum_{n}\mathsf{E}_{\alpha \beta
}^{\left( n\right) }\otimes dA_{t}^{\alpha \beta }$. The corresponding
stochastic derivation is then 
\begin{equation}
dG_{t}=dG\left( \sum_{n}\mathsf{E}_{\alpha \beta }^{\left( n\right) }\right)
.
\end{equation}

\subsubsection{Examples}

i) When the $\mathsf{E}_{11}^{\left( n\right) }=0,$ the relation is simply 
\begin{equation*}
\mathsf{K}=\sum_{n}\mathsf{K}^{\left( n\right) };\quad \mathsf{H}=\sum_{n}%
\mathsf{H}^{\left( n\right) }-\func{Re}\kappa \sum_{\alpha }\mathsf{K}%
^{\left( n\right) \dagger }\mathsf{K}^{\left( n\right) }.
\end{equation*}

\bigskip

ii) (For simplicity, take $\kappa =\frac{1}{2}.)$ Let $\mathsf{W}^{\left(
a\right) },\mathsf{W}^{\left( b\right) }$ be commutative unitaries related
to $\mathsf{E}_{11}^{\left( a\right) }$ and $\mathsf{E}_{11}^{\left(
b\right) }$ by the preceding relations That is, $\mathsf{E}_{11}^{\left(
\alpha \right) }=2i\frac{\mathsf{W}^{\left( \alpha \right) }-1}{\mathsf{W}%
^{\left( \alpha \right) }+1}.$ The composite unitary is $\mathsf{W}=\frac{%
1-i\left( \mathsf{E}_{11}^{\left( a\right) }+\mathsf{E}_{11}^{\left(
b\right) }\right) /2}{1+i\left( \mathsf{E}_{11}^{\left( a\right) }+\mathsf{E}%
_{11}^{\left( b\right) }\right) /2}$ which, after some algebra, becomes 
\begin{equation*}
\mathsf{W}=\mathsf{W}^{\left( a\right) }\frac{\left( 3+\mathsf{W}^{\left(
a\right) }+\mathsf{W}^{\left( b\right) }-\mathsf{W}^{\left( a\right) }%
\mathsf{W}^{\left( b\right) }\right) ^{\dagger }}{\left( 3+\mathsf{W}%
^{\left( a\right) }+\mathsf{W}^{\left( b\right) }-\mathsf{W}^{\left(
a\right) }\mathsf{W}^{\left( b\right) }\right) }\mathsf{W}^{\left( b\right)
}.
\end{equation*}

\bigskip

\bigskip

In \cite{AH}, a formula $K:=L+M+\left[ \left[ L,M\right] \right] $ for the
sum of two stochastic derivations, $L$ and $M$, is given. There the bracket
is $\left[ \left[ L,M\right] \right] $ is their mutual quadratic variation
defined by 
\begin{equation*}
\left[ \left[ L,M\right] \right] \left( t,dt\right) =dL_{t}\circ dM_{t}
\end{equation*}
where $\circ $ denotes composition in $\mathcal{L}\left( \frak{A}\right) $.
We have developed a generalization to arbitrary many summands.


\bigskip

\begin{thebibliography}{99}
\bibitem{S}  R. Stratonovich: \textit{A New Representation For Stochastic
Integrals And Equations} SIAM J.Cont. \textbf{4}, 362-371 (1966)

\bibitem{Sinha}  K.B. Sinha: \textit{Quantum Mechanics of Dissipative Systems%
}, Jour. Ind. Inst. Sciences \textbf{77}, 275-79, (1997)

\bibitem{Weinstein}  A. Weinstein: \textit{The local structure of Poisson
manifolds}\ J. Diff. Geom. \textbf{18}, 523-557 (1985)

\bibitem{Wmod}  A. Weinstein:\ \textit{The modular automorphism group of a
Poisson manifold}, J. Geom. Phys. \textbf{23} , 379-394 (1997)

\bibitem{Meyer}  P.A. Meyer: \textit{A Differential Geometric Formalism for
the It\^{o} Calculus} in Lecture Notes in Mathematics \textbf{851}, Ed. D.
Williams 256-270 (1980)

\bibitem{Emery}  M. Emery: \textit{Stochastic Calculus in Manifolds}
Springer (1980)

\bibitem{JGJMP}  J. Gough, \textit{Dissipative Canonical Flows in Classical
and Quantum Mechanics}, Journ. Math. Phys. \textbf{40}, 2805, (1999)

\bibitem{G}  J. Gough: \textit{Causal structure of quantum stochastic
integrators.} Theoretical and Mathematical Physics \textbf{111}, 2, 218-233
(1997)

\bibitem{ALV}  L. Accardi, Y.G. Lu, I. Volovich: Quantum Theory and its
Stochastic Limit, Springer-Verlag, Berlin, (2002)

\bibitem{G97}  J. Gough, \textit{Noncommumative It\^{o} and Stratonovich
noise and stochastic evolutions}, Theoretical and Mathematical Physics 
\textbf{113}, 2, 1431-1437 (1997), J. Gough: \textit{The Stratonovich
interpretation of quantum stochastic approximations} J. Potential Analysis 
\textbf{11}, 213-233 (1999), J. Gough: \textit{Asymptotic stochastic
transformations for non-linear quantum dynamical systems} Reports Math.
Phys. \textbf{44}, No. 3, 313-338 (1999)

\bibitem{HP1}  R.L. Hudson and K.R. Parthasarathy: \textit{Construction of
quantum diffusions.} Lecture Notes in Math. \textbf{1055}, 384-404 (1984)

\bibitem{H}  A. S. Holevo: \textit{Time-Ordered exponentials in quantum
stochastic calculus}, in Quantum Probability and Related Topics, Vol VII,
pp. 175-202 (1992)

\bibitem{AH}  L. Accardi, R.L. Hudson: \textit{Quantum stochastic flows and
non-abelian cohomology}\ Quantum Probability V, Lecture Notes in Mathematics 
\textbf{1442}, 54-69 (1990)

\bibitem{Sakai}  S. Sakai: \textit{C*-algebras and W*-algebras}\
Springer-Verlag New York (1997)

\bibitem{Lindblad}  G. Lindblad: \textit{On the generators of completely
positive semi-groups} Commun. Math. Phys \textbf{48}, 119-130 (1976)

\bibitem{HP}  R.L. Hudson and K.R. Parthasarathy: \textit{Quantum It\^{o}'s
formula and stochastic evolutions.} Commun.Math.Phys. \textbf{93}, 301-323
(1984)

\bibitem{EVHUD}  M. Evans, R.L. Hudson \textit{Multidimensional Quantum
Diffusions} Quantum Probability III, Eds. L. Accardi and W. von Waldenfels,
Lecture Notes in Mathematics 1303 69-88, (1988)

\bibitem{GQCLT}  J. Gough: \textit{Quantum flows as Markovian limit of
emission, absorption and scattering interactions,} Commun. Math. Phys. 
\textbf{254}, no. 2, 489-512 (2005)

\bibitem{C}  A.N. Chebotarev: \textit{Symmetric form of the
Hudson-Parthasarathy equation} Mat. Zametki, \textbf{60}, 5, 725-750 (1996)

\bibitem{GS05a}  J. Gough, A. Sobolev: \textit{Quantum Markovian
approximations for Fermionic reservoirs}, Inf. Dim. Anal. \& Quantum Prob. 
\textbf{8}, No. 3, 453-471, (2005)
\end{thebibliography}
\end{document}